# Quantum Optics With Single Nanodiamonds Flying Over Gold Films: Towards A Robust Quantum Plasmonics

O. Mollet, A. Drezet and S. Huant

*Institut Néel, CNRS & Université Joseph Fourier, BP 166, F-38042 Grenoble, France*

**Abstract.** A nanodiamond (ND) hosting nitrogen-vacancy (NV) color centers is attached on the apex of an optical tip for near-field microscopy. Its fluorescence is used to launch surface plasmon-polaritons (SPPs) in a thin polycrystalline gold film. It is shown that the quantum nature of the initial source of light is preserved after conversion to SPPs. This opens the way to a deterministic quantum plasmonics, where single SPPs can be injected at well-defined positions in a plasmonic device produced by top-down approaches.

## INTRODUCTION

Entanglement of two qubits by using SPPs[1] would be a starting point for long-distance 2D quantum computing or processing with SPPs fully compatible with integrated-photonics technology. In this context, we use here a quantum source of light made of a single fluorescent ND to launch SPPs in a polycrystalline gold film and analyze the quantum nature of the leaky plasmonic light. We find that the quantum nature of the initial light is preserved after conversion to SPPs.

## SURFACE PLASMON LAUNCHING WITH A NANODIAMOND-BASED OPTICAL TIP

Using an all-optical scheme centered on a near-field scanning optical microscope (NSOM), we can reproducibly attach a single ≈25 nm ND hosting a few (1 to 5) NV centers onto the apex of an optical fiber-tip. This achieves single-ND-based active optical tips. We have implemented such tips in a NSOM environment. Illuminating the grafted ND with a laser light (515 nm wavelength) guided by the fiber itself and using the fluorescence light generated by the ND as source of light achieves a genuine scanning single-photon microscopy.[2] This microscopy operates at room temperature and on a long term thanks to the exceptional photostability of the NVs.

Furthermore, we have shown that a ND-based tip efficiently launches SPPs into gold films when placed in the near field (ND-gold distance << 100 nm): see Fig.1.[3] This works because the ND fluorescence has the right energy and the evanescent waves forming the near field have large spatial frequencies able to transfer enough momentum to SPPs.

Since the ND is a quantum source of light, a limited number of SPPs, depending on the actual NV occupancy, contributes to the experimental images. A critical prerequisite for using these SPPs in future quantum protocols is that the plasmon conversion and propagation along the metal film do not destroy the second-order coherence of the initial source of light. This issue is addressed in this paper.

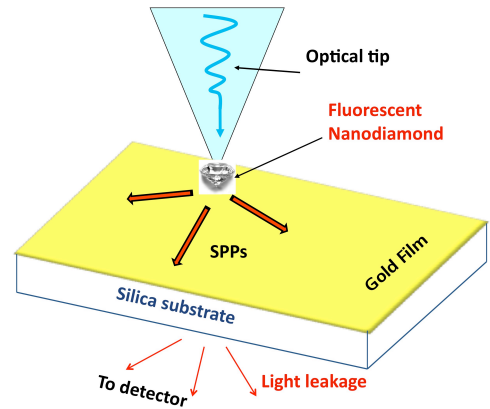

**FIGURE 1.** Principle of the SPP near-field launching by a ND-based tip. A single ND is glued at the apex of an optical tip. The fluorescence of the NVs is coupled to SPPs propagating along a gold film. Light transmitted trough the film and leaking into the substrate is collected by an objective (not shown) for subsequent imaging or analysis.

## SECOND-ORDER COHERENCE OF THE PLASMONIC LIGHT

From previous works, it is known that a characteristic signature of SPP propagation along a metal film is obtained by leakage radiation microscopy

(LRM) through that film, provided the film thickness (here 30 nm) is not too large to damp the plasmonic field.[3-6] LRM uses the phase matching of the plasmonic field (conservation of the in-plane momentum) at the metal-dielectric substrate (here silica) that takes place at angles above the critical incidence for total internal reflection in this dielectric, i.e., in the so-called forbidden light region. Then, collecting this plasmonic light with a high numerical aperture objective (>1) leads to a blunt optical signal extending over the SPP propagation length ($\approx$ 3 µm here) in the direct space and to a characteristic circle in the Fourier space. Both the real-space and Fourier signals can be imaged directly on a camera[3,7] or be used in photon-correlation measurements by injecting them in a Hanbury-Brown Twiss correlator.[8]

From the correlation measurements, we can extract the time-intensity second-order correlation function $g^2(\tau)$, i.e., the normalized conditional probability of detecting a second photon at time $\tau$ when one photon has already been detected at zero time. For $N$ equivalent emitters, we have:[9]

$$g^2(\tau) = 1 - \frac{\rho^2}{N} e^{-\gamma\tau}. \quad (1)$$

In (1), $\gamma$ is linked to the pumping rate $R$ and emitter lifetime $T$: $\gamma = R+1/T$, and $\rho=S/(S+B)$, with $S$ the NV fluorescence signal and $B$ the incoherent background (e.g. gold fluorescence). From the fit to $g^2(\tau)$ (see below) and the measured saturation behavior of the emitter power, we can extract all relevant parameters, including $N$.[10]

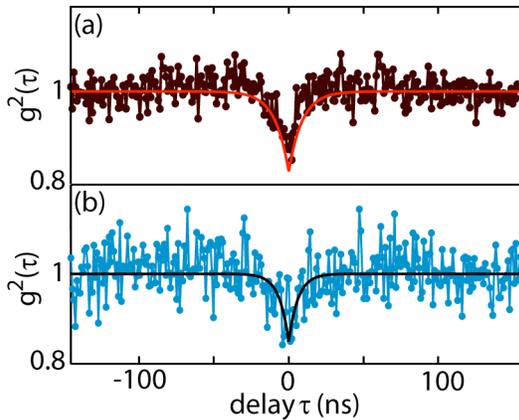

**FIGURE 2.** Antibunching behavior of a 5-NV based tip (a) and of the plasmonic light collected by LRM after SPP launching in a 30 nm gold film (b). Full lines are fit to the data as explained in the text.

Figs. 2(a) and (b) show the measured and fitted $g^2(\tau)$ for a ND-based tip facing the dielectric substrate and the gold film, respectively. In the first case, there is no SPP contribution at all and the correlation trace is that of the initial source of light. It exhibits a clear antibunching dip at zero delay, from which we find that the involved number of emitters (NV centers) is $N= 5$. The analysis of the second trace, taken in the Fourier space by LRM through the gold film, also gives $N= 5$ emitters, with a marginally decreased $T$ from 9 ns to 6 ns. This decrease is possibly due to the self-interaction of the emitters close to the metal.[11]

## CONCLUSION

The above discussion shows that the second-order coherence of a quantum source of light is preserved after SPP injection into a polycrystalline film. Such a film can be patterned by standard lithography techniques so that quantum plasmonic devices, where the quantum emitter(s), waveguides, and other necessary components[12] would all be integrated on a single chip can be envisioned.[10]

## ACKNOWLEDGMENTS

We acknowledge support from the French ANR through the NAPHO and PLASTIPS projects.